\documentclass{agujournal}
\journalname{Space Weather}
\graphicspath{ {Figures/}}
\begin{document}

\title{Highly relativistic electron flux enhancement during the weak geomagnetic storm of April--May 2017}

\authors{Ch. Katsavrias\affil{1,2}, I. Sandberg\affil{3}, W. Li\affil{4}, O. Podladchikova\affil{5}, I.A. Daglis\affil{1,2,6}, C. Papadimitriou\affil{3}, C. Tsironis\affil{2,7} and S. Aminalragia-Giamini\affil{3}}

\affiliation{1}{Department of Physics, National and Kapodistrian University of Athens, Greece.}
\affiliation{2}{Institute of Accelerating Systems and Applications, National and Kapodistrian University of Athens, Greece.}
\affiliation{3}{Space Applications and Research Consultancy (SPARC), Athens, Greece.}
\affiliation{4}{Center for Space Physics, Boston University, USA.}
\affiliation{5}{Solar--Terrestrial center of excellence, Royal Observatory of Belgium.}
\affiliation{6}{Institute of Astronomy, Astrophysics, Space Applications and Remote Sensing, National Observatory of Athens, Greece.}
\affiliation{7}{School of Electrical and Computer Engineering, National Technical University of Athens, 157 73 Athens, Greece.}
\correspondingauthor{Ch. Katsavrias}{ckatsavrias@phys.uoa.gr}

%  List up to three key points (at least one is required)
%  Key Points summarize the main points and conclusions of the article
%  Each must be 100 characters or less with no special characters or punctuation 

\begin{keypoints}
\item Long--lasting multi--MeV electron enhancement during a period of a relatively weak geomagnetic storm not recorded in GEO.

\item Electron seed population was accelerated to relativistic energies by the enhanced chorus waves.

\item Relativistic electrons were further accelerated up to 10 MeV by inward diffusion
\end{keypoints}

\begin{abstract}
We report observations of energetic electron flux and Phase Space Density (PSD) to show that a relatively weak magnetic storm with $Sym-H_{min} \approx -50 nT$, resulted in a relativistic and ultra--relativistic electron enhancement of two orders of magnitude similar to the St. Patrick's event of 2015, an extreme storm with $Sym-H_{min} \approx -235 nT$. This enhancement appeared at energies up to $\approx 10$ MeV, lasted for at least 24 days and was not recorded in geosynchronous orbit where most space weather alert data are collected. By combined analysis of PSD radial profiles and Fokker--Planck simulation, we show that the enhancement of relativistic and ultra--relativistic electrons is caused by different mechanisms: first, chorus waves during the intense substorm injections of April 21--25 accelerate the seed electron population to relativistic energies and redistribute them while inward diffusion driven by Pc5 ULF waves further accelerates them to ultra--relativistic energies.
\end{abstract}
% ------------------------------------------------------------------------ %
%  TEXT
% ------------------------------------------------------------------------ %
%% Suggested section heads:
% \section{Introduction}
% The main text should start with an introduction. Except for short
% manuscripts (such as comments and replies), the text should be divided into sections, each with its own heading. 
% Headings should be sentence fragments and do not begin with a  lowercase letter or number. Examples of good headings are:
% \section{Materials and Methods} % Here is text on Materials and Methods.
% \subsection{A descriptive heading about methods} % More about Methods.
% \section{Data} (Or section title might be a descriptive heading about data)
% \section{Results} (Or section title might be a descriptive heading about the results)
% \section{Conclusions}
% ------------------------------------------------------------------------ %

\section{Introduction}

Earth's outer radiation belt response to geospace disturbances is extremely variable.  Previous studies have shown that the trapped relativistic electron population, in this near--Earth space, can be enhanced, depleted, or even not affected at all \citep{Reeves2003,Turner2013,Reeves2016} due to the interplay of acceleration and loss mechanisms.

Concerning the acceleration process we can identify two important mechanisms which rely on "radial diffusion" and local acceleration \citep{Balasis2016}. For inward radial diffusion, a "reservoir" of electron phase space density (PSD) in the plasma sheet is required and, with the breaking of the third adiabatic invariant, these particles  diffuse inwards in radial distance and gain energy due to the conservation of the first adiabatic invariant \citep{Taylor2004}. A positive gradient in the PSD radial profile is the condition leading to this inward diffusion. For local acceleration, substorm injections can provide the outer radiation belt with a seed population of tens to a few hundreds keV electrons which then are \textit{in situ} accelerated to MeV energies by violating their first and second invariants \citep{Horne2005,Meredith2003,Thorne2013,Bortnik2016}. Unlike radial transport, "local acceleration" produces a rising peak in PSD where the resonant interaction occurs. During most geomagnetic disturbances these two mechanisms act together making it difficult to discriminate each drivers' effect. Recently, \citet{Li2016} showed that radial diffusion alone underestimates the observed electron acceleration during St. Patrick's event of 2015 being unable to produce the rising peak at the right location. Yet, argued that further electron acceleration, to even higher energies, could be achieved by radial diffusion via Pc5--ULF waves. The energy--dependent radial extent of the electrons in the outer radiation belt was discussed in detail by \citet{Ma2016}. By examining the March 1, 2013 storm, they showed that both local acceleration by chorus and radial diffusion by ULF waves were needed to reproduce the observed electron flux evolution of 0.59 and 3.4 MeV. To the same extent, \citet{Jaynes2015} proposed that ULF waves, through enhanced inward radial diffusion of  relativistic electrons in the 1-2 MeV energy range, can lead to enhancement of several MeV electron fluxes. Moreover, \citet{Liu2018}, examined the 2--3 October 2013 event and indicated that the observed enhancement of electrons in the 1.8--3.4 MeV energy range occurred due to radial diffusion of the already accelerated seed electrons from L$\approx$8.3.

The correlation of geomagnetic activity with the enhancements of MeV electron flux in the outer radiation belt has been recognized for decades \citep{Reeves2003,Kataoka2006,Baker2007}. Many recent statistical studies have used a threshold of Dst (or Sym-H index) as a selection criterion for events of geospace disturbances \citep{Murphy2017,Moya2017}. 

On the other hand, recent studies have argued that Dst index is a poor indicator of storm effects because it potentially neglects acceleration mechanisms that cause the enhancement of radiation belt electrons during relatively quiet geomagnetic conditions \citep{Borovsky2006}. In addition, \citet{Schiller2014} showed that during a period of weak geomagnetic disturbances ($Dst_{min} \approx -30 nT$) there was an enhancement of approximately 2.5 orders of magnitude for 0.6 -- 1.3 MeV electrons in less than 13 h. Along the same line, \citet{Katsavrias2015a} showed that during a period of continuously positive Sym--H index, an electron PSD dropout of 2 orders of magnitude occured for electrons with $\mu>300$ MeV/G. 

In this work we present a detailed analysis of the events that occurred during the period April 16 -- May 16, 2017 when a long--lasting enhancement of relativistic and ultra--relativistic electrons occurred during a relatively weak geomagnetic storm. Events such as the one described in this letter demonstrate that the classification of geomagnetic activity using geomagnetic indices (such as Dst) can exclude some significant enhancement events.

\section{Data Selection and Methodology}\label{Data}

We use the high--quality pitch angle distributions from the Magnetic Electron Ion Spectrometer--MagEIS \citep{Blake2013}, and the Relativistic Electron Proton Telescope--REPT \citep{Baker2012} on--board the Van Allen Probes (from now on RBSP) in order to calculate PSD for fixed values of the three adiabatic invariants using the method described in \citet{Chen2005}. Values of the K and $L^{*}$ invariants are provided from the magnetic ephemeris of the ECT--suite and are calculated at each measurement point using the \citet{Tsyganenko2005} magnetospheric field model (TS05). In addition, we use magnetic field measurements from the fluxgate magnetometers of RBSP \citep{Kletzing2013} with 3 minutes resolution. By applying wavelet analysis, using the method described in \citep{Balasis2013}, we calculate the weighted sum of the wavelet spectrum over Pc5 wave frequencies (typically between 2 and 7 mHz). Finally, we use the recently developed technique by \citet{Li2013} to infer lower--band chorus wave amplitudes. Supplementary measurements of five minutes averaged values of solar wind speed, dynamic pressure and interplanetary magnetic field mapped at 1 AU as well as geomagnetic indices Sym-H and AL from the NASA/OMNI database are also considered.

\section{Results and Discussion}\label{sec:disc}

%------------------------------------------------------------------------------------
 \begin{figure}[h]
 \centering
 \includegraphics[width=25pc]{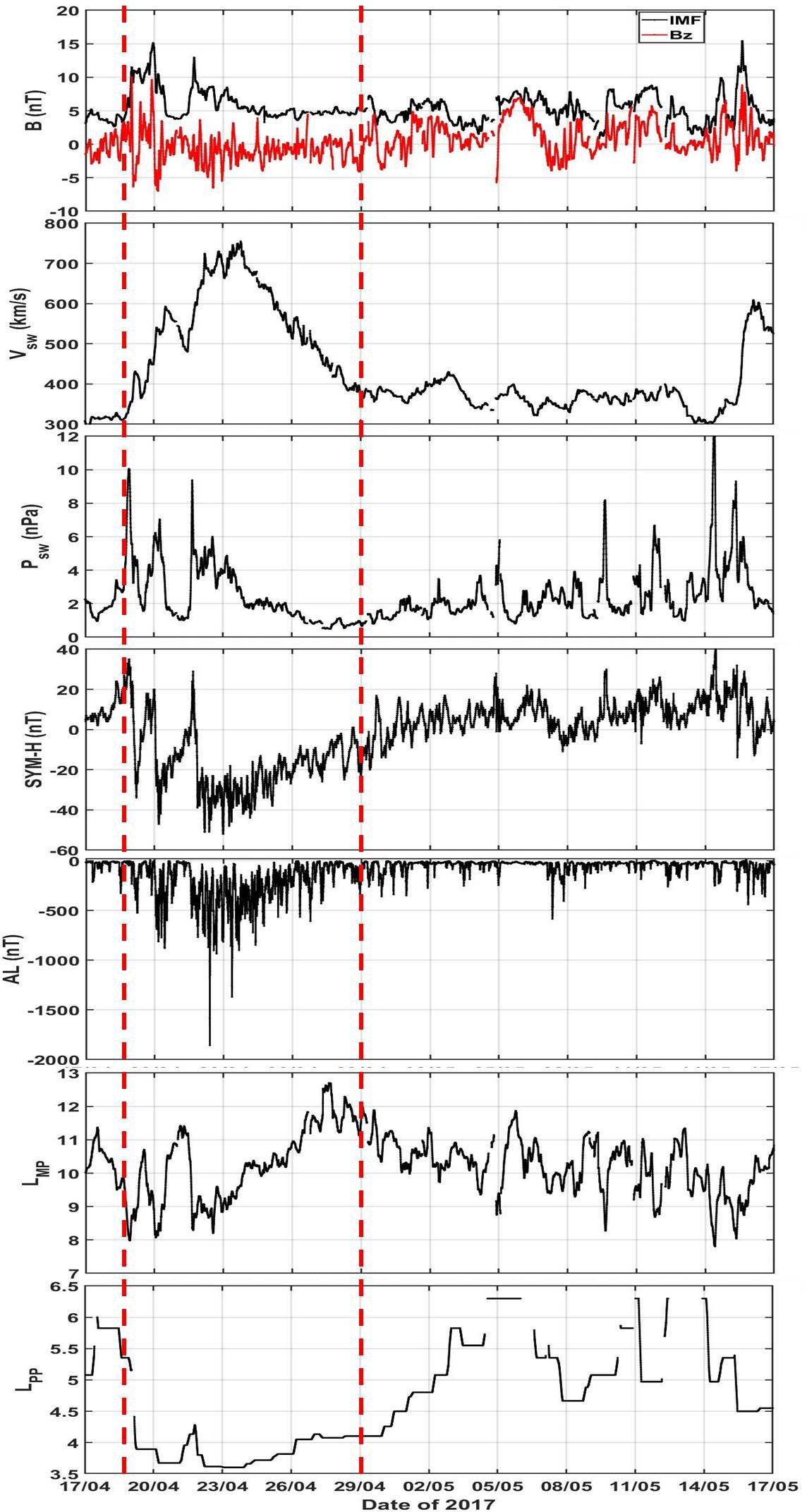}
 \caption{Overview of the April 17 -- May 16, 2017 solar wind and magnetosphere conditions with 5 minutes resolution. Top to bottom: Interplanetary Magnetic Field (IMF) and its z--component ($B_z$), Solar Wind speed, dynamic pressure, AL index, SYM-H index, dayside magnetopause location \citet{Shue1998} and MLT averaged plasmapause location based on the model of \citet{O'Brien2003b}. The vertical red dashed lines indicate the beginning and the end of the geospace disturbances.}
 \label{IMF}
  \end{figure}
%------------------------------------------------------------------------------------

An overview of the solar wind and magnetosphere conditions during April 17 -- May 16, 2017 is presented in figure \ref{IMF}. After a short--lived, pulse from an ICME which arrived at 1 AU on April 18, a high speed stream from an equatorial (positive polarity) coronal hole arrived to the Earth on April 19, according to measurements by ACE and DSCOVR satellites reported by SIDC weekly bulletin on solar and geomagnetic activity. The solar wind speed raised to 600 km/s (which is conventionally considered as the lower limit of the fast solar wind speed) with interplanetary magnetic field of 15 nT and its Bz component of -6 nT. On April 21, ACE and DSCOVR, provided signatures of the arrival of a high speed stream from another (negative) coronal hole with speed reaching 800 km/s and interplanetary magnetic field of 12 nT. The influence of this high speed stream continued until April 23, while on Monday April 24 the Earth was still inside a fast solar wind $\approx720$ km/s. The solar wind speed was slowly decreasing and later (April 27) the Earth was inside the slow solar wind. 

These three events produced three pressure pulses on April 18 (10 nPa at 19:00 UT), April 20 (8 nPa at 05:00 UT) and April 21 (15 nPa at 16:00 UT), combined with three consecutive magnetopause compressions (to $L \approx 7.5$) followed by three decreases in the Sym-H index with minimum at -35, -40 and -50 nT respectively. The plasmapause -- located at $L \approx 6$ before the CME -- was compressed to 3.5--4 until April 30. Finally, a series of intense substorm activity (as indicated by the AL index) occurred during April 19 and April 21 to 26 with minimum AL at $\approx -1800$ nT on April 22. When the Earth was under the influence of this high speed stream, GOES detected flux enhancements for electrons with energies higher than 0.8 and 2 MeV (see figure S1 in supplementary material). We define as the end of the geospace disturbances, the April 29, as speed was decreased to values below 400 km/s and the Sym-H index was continuously above -20 nT. Also note, that all solar wind parameters and geomagnetic indices remained relatively weak until May 14, when another driver affected the Earth as indicated by the increased solar wind speed and pressure.

%------------------------------------------------------------------------------------
 \begin{figure}[h]
 \centering
 \includegraphics[width=25pc]{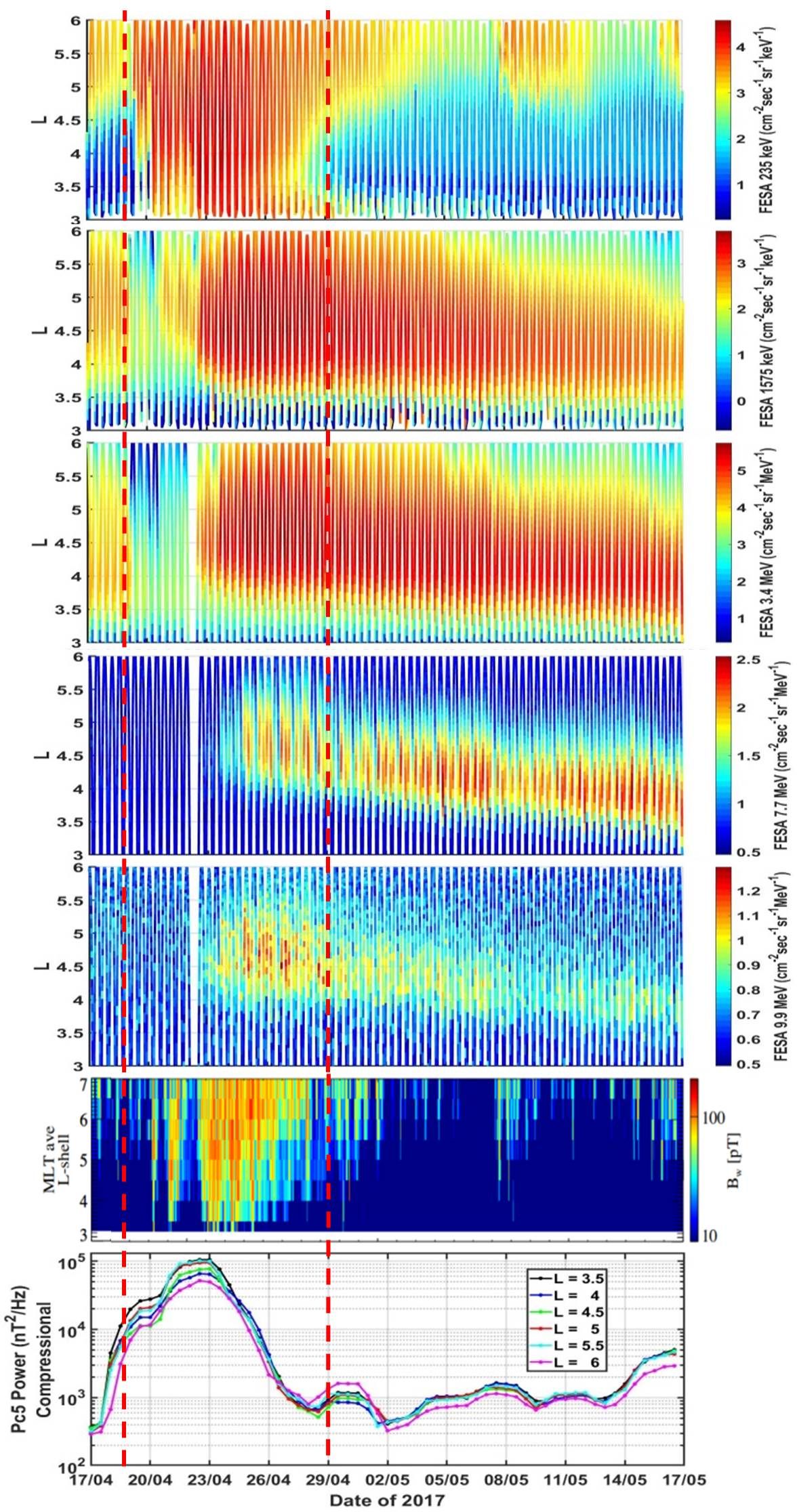}
 \caption{Spin--averaged electron differential fluxes measured by the MagEIS and REPT instruments on--board RBSP A--B throughout the outer belt along with chorus amplitudes and averaged power for compresional Pc5 waves. Top to bottom: fluxes for 235, 1575, 3400, 7700 and 9900 keV, chorus amplitudes and averaged power for compressional Pc5 waves. The colorbar represents the log values of Spin--averaged electron differential fluxes. The vertical dashed lines indicate the beginning and the end of the geospace disturbances.}
 \label{FLUX}
  \end{figure}
%------------------------------------------------------------------------------------

Variations of spin--averaged electron flux during the considered time period are shown in figure \ref{FLUX}. The seed electron population (235 keV) begun to enhance on April 19 at L>4. From April 20 until, approximately, April 29 the seed population was enhanced by 3 orders of magnitude at all L--shells while remaining enhancements at L>4.5 occurred up to May 3. These enhancements coincide with the beginning and the end of the substorm activity. Moreover, a weaker enhancement at L>5 occurred during May 8--11. Electrons with energies of few MeV showed a flux dropout of approximately two orders of magnitude which coincides with the maximum compression of the magnetopause (April 18) and then began to accelerate fairly rapidly at lower energies (1575 keV), followed by acceleration at even higher energies but with a time delay (3.4 MeV are increased on April 23). The ultra--relativistic population (7.7 MeV) shows a flux enhancement of about 2 orders of magnitude limited around $L \approx 4 - 5$. This enhancement began on late April 23 and lasted at least 24 days. After April 29, the peak location of ultra--relativistic electron flux gradually moved to lower L--shells ($L \approx 4 - 4.5$) while after May 8 reached L=3.5--4.5. A quite similar evolution was  observed for the 9.9 MeV electron flux but the enhancement was up to 1 order of magnitude. It should be noted, that the 9.9 MeV energy channel is the last REPT channel that shows flux enhancement. 

Here we emphasize the fact that a weak geomagnetic storm with $Sym-H_{min} \approx -50 nT$, resulted in a relativistic and ultra--relativistic electron enhancement of 2 orders of magnitude similar to the St. Patrick's event of 2015, the strongest storm ($Sym-H_{min} \approx -235 nT$) seen over the past decade \citep{Baker2016}. This enhancement lasted for at least 24 days and appears to energies up to 10 MeV as measured by RBSP--A/REPT or RBSP--B/REPT. Moreover, the importance of this event lies in the fact that the enhancement of electron flux larger than 6 MeV was not recorded at geosynchronous orbit (as in St. Patrick's event) where most space weather monitoring data, which are used for issuing alerts, are collected (also see figure S1 in supplementary material).

Whistler mode chorus waves -- following the series of intense substorms -- were enhanced during the time period 20--27 April at all L--shells while there was a remaining amplitude enhancement at L > 5 up to May 1st (figure \ref{FLUX}) and moreover, a short--lived, intermittent activity at L > 5 during 7--8 May. This remarkable consistency of chorus and substorm activity (as indicated by the AL index) is in agreement with the results of  \citet{Meredith2001,Meredith2002} and \citet{Boynton2018} who showed that there is a high correlation between chorus waves activity and substorm injections. Compressional Pc5 ULF waves, on the other hand, appeared with pronounced power at all L--shells (maximum power at $5 \leq L \leq 5.5$) during the whole April 18--26 time period following the high values of solar wind speed. Smaller and intermittent enhancements of Pc5 power occurred even after the aforementioned period and up to May 12 probably due to the small variations in solar wind pressure. These results are consistent with the statistical study of \citet{Liu2010} over a 21--month period who showed that ULF wave activity (Pc4--5 frequency range) at 4--9 R$_E$ was strongly correlated with solar wind speed and pressure.  
%------------------------------------------------------------------------------------
 \begin{figure}[h]
 \centering
 \includegraphics[width=30pc]{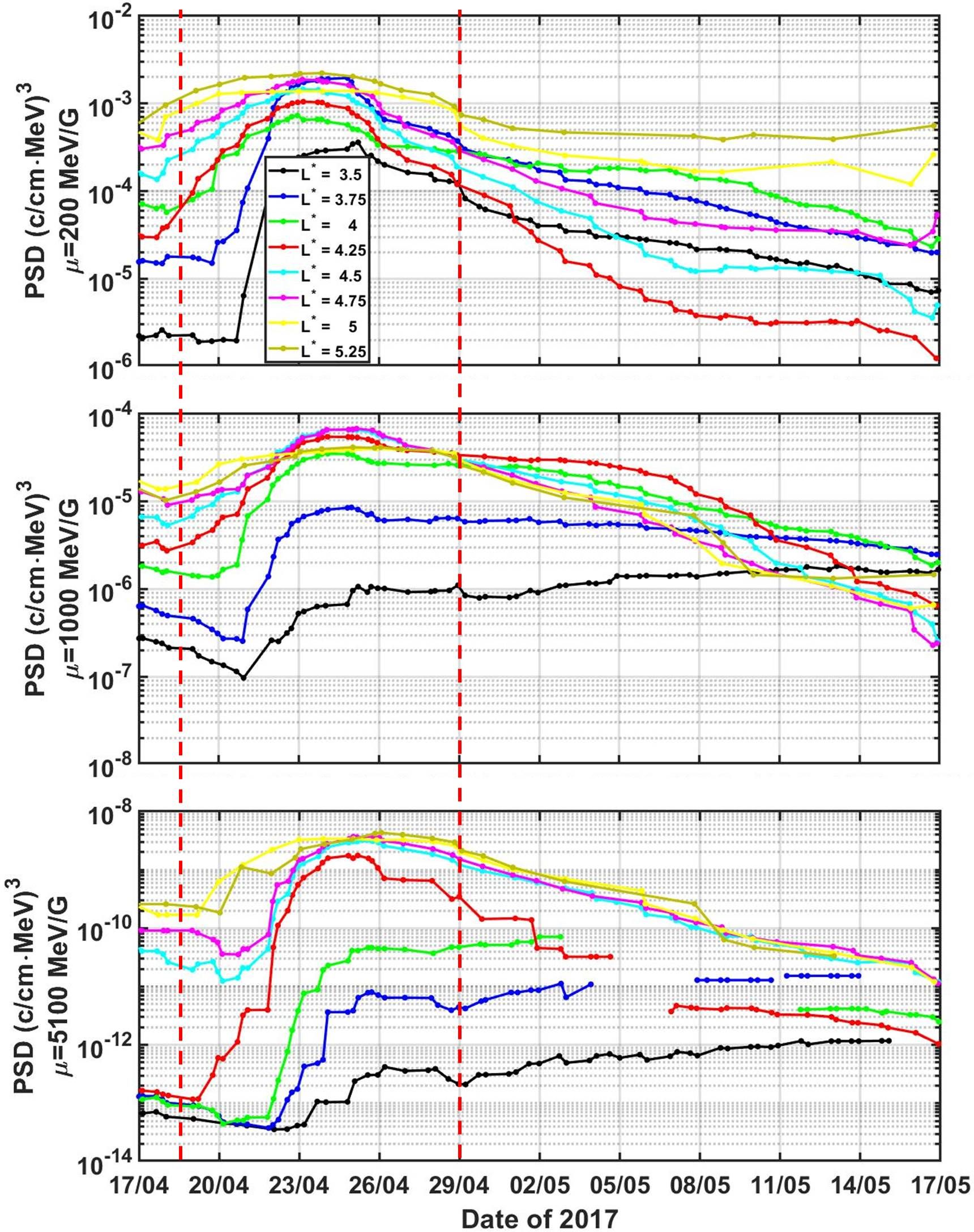}
 \caption{PSD time profiles of near--equatorial mirroring electrons (K$\le0.03$ G$^{1/2}$R$_{E}$) for fixed values of the three adiabatic invariants during the April 17 -- May 16 time period. Each panel corresponds to different value of $\mu$ (top to bottom: 200, 1000 and 5100 MeV/G, respectively) and each color curve to different value of L$^{*}$. The vertical dashed lines indicate the beginning and the end of the geospace disturbances.}
 \label{timeprofiles}
  \end{figure}
%------------------------------------------------------------------------------------

Time evolution of PSD for fixed values of the three adiabatic invariants is shown in figure \ref{timeprofiles}. Each point corresponds to the average PSD (with approximately 8 hours resolution) calculated from both RBSP A and B for near--equatorial mirroring electrons with K$\le0.03$ G$^{1/2}$R$_{E}$. The latter corresponds to equatorial pitch angle values (figure S2 in supplementary material) varying between 70 to 80 degrees during April 17--20, 80--90 degrees during April 21--27 and 60 to 80 degrees during April 28--May 16 at the heart of the outer belt (L$^{*}=4.5$). Moreover, we have chosen 3 values of the first adiabatic invariant, namely 200, 1000 and 5100 MeV/G, which, at $4\leq L^{*}\leq5$, roughly correspond to energies of approximately 0.2--0.40, 1--2 and 3--5 MeV (figure S3 in supplementary material). Seed population with $\mu$ = 200 MeV/G, starts to increase shortly before April 19 at L$^{*}\geq4.5$. Subsequently, following the AL decrease (AL$_{min}\approx -1000$ nT at 10:00 UT on April 20 and AL$_{min}\approx -2000$ nT at 09:30 UT on April 22), the seed population was enhanced at lower L--shells on April 20 (L$^{*}<4.5$). By April 29 (end of the substorm series), seed population was decreased at pre--event levels (or less) for L$^{*}>4$ while the inner edge population (L$^{*}<4$) remained enhanced. Relativistic population with 1000 MeV/G, shows enhancements of approximately 1 order of magnitude beginning on April 20 at higher L--shells which, gradually decreases after April 29 (maximum on April 24). At L$^{*}<4$, PSD exhibits a sudden decrease and then an up to 2 orders of magnitude enhancement which remained enhanced until the end of the examined time period at L$^{*}<4.5$. Electron population with 5100 MeV/G shows a PSD dropout during 20--21 April at higher L$^{*}$, and then a sudden increase (approximately 4 orders of magnitude at L$^{*}$=4.25) on late April 22. PSD at higher L$^{*}$ began a slow decrease after the peak on April 25 while PSD in lower L$^{*}$ continued to increase.

%------------------------------------------------------------------------------------
 \begin{figure}[h]
 \centering
 \includegraphics[width=30pc]{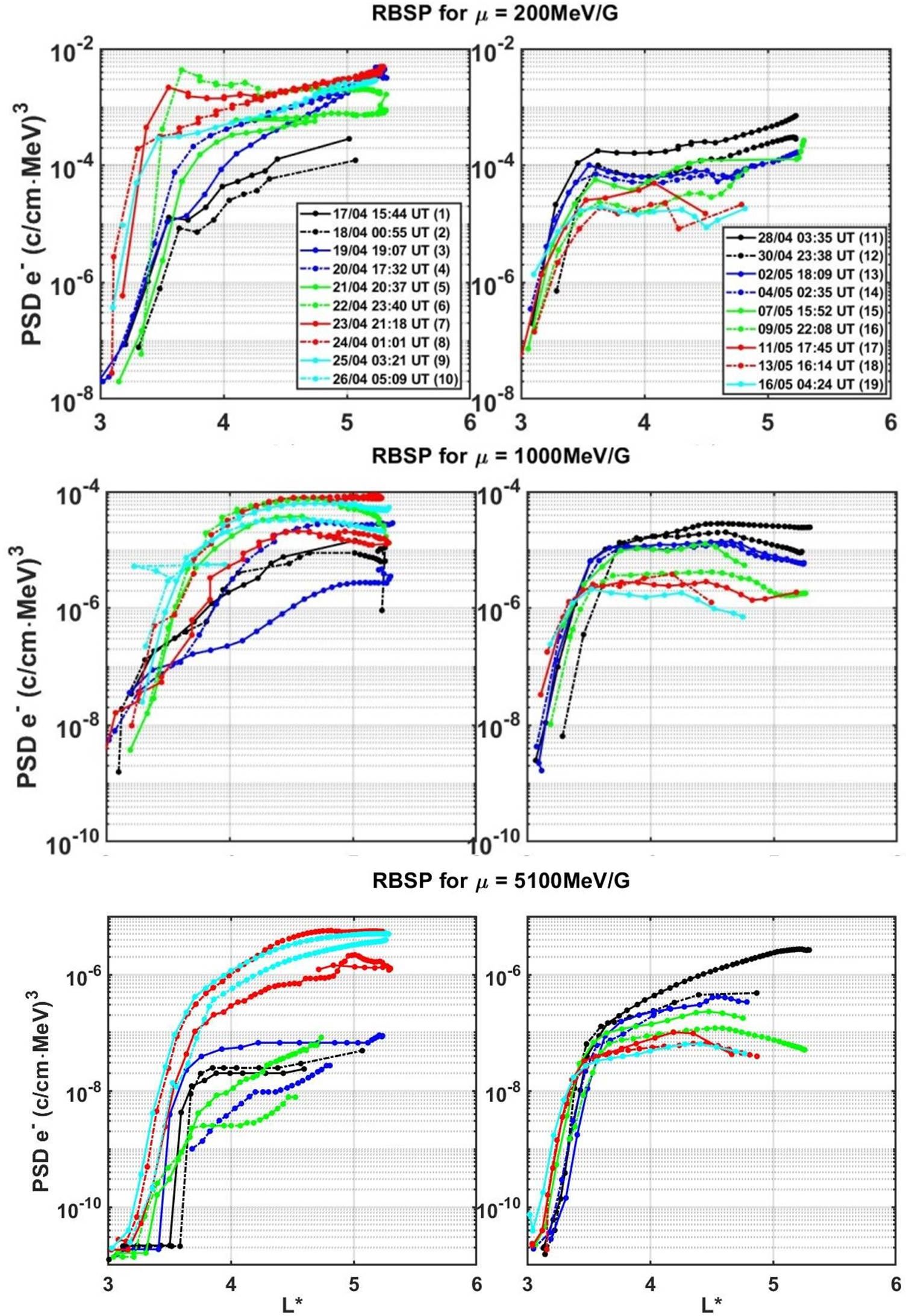}
 \caption{Time evolution of near--equatorial mirroring electron PSD as a distribution of $L^{*}$ for 3 fixed values of $\mu$ (200, 1000 and 5100 MeV/G) and K$\le0.03$ G$^{1/2}$R$_{E}$ for the April 17--May 16, 2017 time period.}
 \label{PSDvsL}
  \end{figure}
%------------------------------------------------------------------------------------

Furthermore, we calculated electron PSD as a function of L$^{*}$ for fixed values of $\mu$ and K invariants in order to investigate non--adiabatic electron dynamics. The results are shown in figure \ref{PSDvsL} from the beginning of April 17 to the end of May 16, 2017. Electron PSD with $\mu=200$ MeV/G was enhanced right after the beginning of the substorm activity. In detail, there was an increase at L$^{*}>4$ on April 19 (orbit 3) following an AL$_{min}\approx -600$ nT at 06:00 UT on April 19. On April 20, PSD was increased at L$^{*}<4$ as well following an AL$_{min}\approx -1000$ nT at 10:00 UT. Subsequently, following an AL$_{min}\approx -2000$ nT at 09:30 UT on April 22, PSD was further increased at lower L--shells as a consequence of the penetration of substorm injected electrons deep into the inner magnetosphere. After April 26, when substorm activity had faded, PSD returned to pre--event levels only for L$^{*}>4$ while at lower L--shells (L$^{*}\approx3.5$) PSD remained enhanced at 2 orders of magnitude. The continuously flat/positive PSD gradient suggests that these seed electrons were primarily transported from higher to lower L$^{*}$. The electron PSD for 1000 MeV/G , shows a sudden depletion (April 19 -- orbit 3) and then was increased up to 2 orders of magnitude exhibiting rising peaks at L$^{*}\approx4.5$ from April 21 until April 25 (orbits 5--9). These rising peaks, which indicate local heating processes, coincide with the occurrence of intense substorm activity and the enhancement of chorus amplitude. This result is consistent with the findings of \citet{Katsavrias2015b} who showed that, during the March 2013 intense storm and concerning the $>300$ MeV/G electrons, the chorus--driven acceleration exceeded the Pc5--driven losses. After April 26, PSD gradients became flat and there was a gradual depletion down to pre--event levels. Note, that this gradual depletion starts from April 28, lasts up to May 15 and coincides with the expansion of the plasmapause. The latter suggests that plasmaspheric hiss, which is limited mostly inside the plasmapause \citep{Thorne2005} and is able to precipitate electrons with a timescale from $\approx1$ day to tens of days depending on energy \citep{Jaynes2014}, could be the reason of that depletion. The electron PSD for 5100 MeV/G, shows a 3 orders of magnitude enhancement starting from late April 23 at all L--shells (orbit 7) which seems to return to pre--event levels after May 9 (orbit 16). These ultra--relativistic electrons -- in contrast with the relativistic ones -- show continuously positive gradients throughout the whole period of interest, indicating that inward radial diffusion could be the dominant mechanism for this acceleration. Nevertheless, we cannot exclude the possibility that local acceleration, could occur at L$^{*}$>5 \citep{Schiller2014,Boyd2018} where RBSP's limited radial coverage could not observe. The latter could also produce the observed positive gradient in PSD.

In order to provide additional evidences for the aforementioned results, we solve the Fokker--Planck equation (with radial diffusion only), for the electron population with $\mu$ 1000 and 5100 MeV/G and K$\le0.03$ G$^{1/2}$R$_{E}$. 

\begin{equation}
\frac{\vartheta f}{\vartheta t} = L^{2} \frac{\vartheta}{\vartheta L}(\frac{D_{LL}}{L^{2}}\frac{\vartheta f}{\vartheta L})- \frac{f}{\tau}
\label{FokkerPlanck_diff}
\end{equation}

For the calculation of the D$_{LL}$, instead of \textit{in situ} measurements, we use a Kp-- and $\mu$--dependent radial diffusion coefficient. In detail, following \citet{Ma2016}, we have used D$^{E}_{LL}$ from \citet{Liu2016}:

\begin{equation}
D_{LL}^E <Liu> = 1.115 \cdot {10^{ - 6}} \cdot {L^{8.184}} \cdot {\mu ^{ - 0.608}} \cdot {10^{0.281 \cdot {K_P}}}
\label{DLLE}
\end{equation}

and D$^{B}_{LL}$ from \citet{Ozeke2014}:

\begin{equation}
D_{LL}^B <Ozeke> = 6.62 \cdot {10^{ - 13}} \cdot {L^8} \cdot {10^{ - 0.0327 \cdot {L^2} + 0.625 \cdot L - 0.0108 \cdot {K_P}^2 + 0.499 \cdot {K_P}}}
\label{DLLB}
\end{equation}

The selection of this combination, instead of the classical approach of \citet{Brautigam2000} \citep{Zhao2018}, can be justified as follows; the latter overestimates the diffusion coefficient at L>4, especially at higher values of $\mu$. In addition, $D_{LL}^E <Liu>$ contains an energy dependency which is not present in D$_{LL}$  by both \citet{Ozeke2014} and \citet{Brautigam2000}. Nevertheless, this combination exhibits large errors for $\mu$ values outside  the 400--8000 MeV/G range \citep{Liu2016} and, thus, we only perform simulation for $\mu$=1000 and 5100 MeV/G. 

The numerical simulation has been performed for the period of April 19--29. During this period, the plasmapause is approximately located at L$\leq4$, allowing us to exclude further loss due to pitch angle scattering from plasmaspheric hiss by setting tau =$\infty$. The inner and outer boundaries are set at L$^{*}$=3.2 and 5.2, respectively, in order to have, the least amount of data gaps and thus, interpolated values.

%------------------------------------------------------------------------------------
 \begin{figure}[h]
 \centering
 \includegraphics[width=30pc]{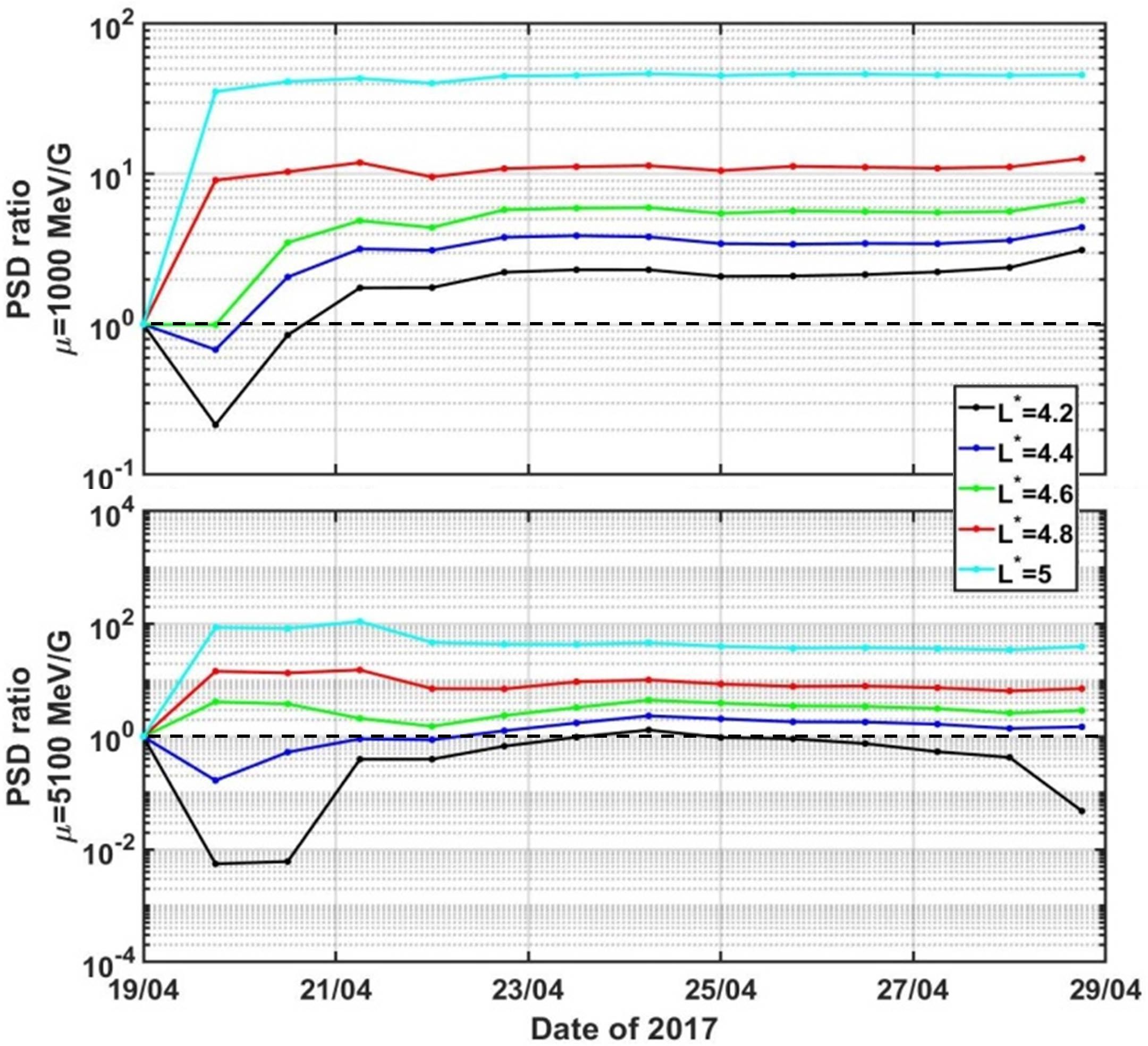}
 \caption{Time evolution of the ratio of electron PSD from RBSP (PSD$_{RBSP}$) over the simulated electron PSD (PSD$_{SIM}$) at various L$^{*}$ for three values of $\mu$ (top to bottom: 1000 and 5100 MeV/G) and K$\le0.03$ G$^{1/2}$R$_{E}$ during the April 19--29 time period. The horizontal dashed lines, correspond to 1, where PSD$_{RBSP}$=PSD$_{SIM}$.}
 \label{PSDratio}
  \end{figure}
%------------------------------------------------------------------------------------

Figure \ref{PSDratio} shows the ratio of near--equatorial electron PSD, calculated from RBSP measurements, over the simulated electron PSD from our model at various L$^{*}$. Note, the average location of the plasmapause, during the time--period of interest, is at approximately L=3.8. Since, there is a very good correlation of the plasmapause location and the maximum penetration of ULF waves \citep{Georgiou2015}, which in return can cause radial diffusion, we do not show results at L$^{*}\leq$ 3.8. As shown, there are large with respect to 1 value during April 19--21 for all $\mu$ values and at all L$^{*}$. A possible reason could be the fast and strong variations of both the plasmapause and the magnetopause location on April 19--21 which, of course, the model cannot reproduce. After April 23, the  electron population with $\mu$=5100 MeV/G is the one best simulated by our model with PSD ratio very close to 1, especially at the heart of the outer belt at $4.2\leq L^{*} \leq 4.8$. The latter -- combined with the continuously positive gradients -- suggests that inward diffusion process driven by the enhanced Pc5 activity, during 19--29 April, is the reason for the acceleration of electrons higher than 3 MeV. This scenario is also supported by the power spectral density of the observed compressional Pc5 waves. As shown in figure S4 (see supporting material), the peak in the power spectral density is, roughly, in the 2--3.5 mHz frequency range. The latter includes the 2.3--3 mHz frequency range, which corresponds to the drift frequency of near--equatorial electrons with $\mu=$1000 MeV/G.

All of the above, support the scenario that the enhancement of electrons with E>3 MeV is caused by the inward radial diffusion of the 1--2 MeV electrons, which, in turn, were accelerated via local heating due to the enhanced chorus activity.

\section{Conclusions}	

The combined analysis of PSD radial profiles and radial diffusion simulation of the mid--April to mid--May, 2017 period has shown that:

\begin{enumerate}
	\item Seed electrons were enhanced during the April 18--29 following the arrival of a high--speed solar wind stream and the ensuing series of intense substorms and chorus activity throughout the whole outer radiation belt.
	\item This seed population was accelerated to relativistic energies by the enhanced chorus waves during the April 21--25 time period exhibiting characteristic rising peaks in the PSD radial distribution. 
	\item Finally, these relativistic electrons were further accelerated up to ultra--relativistic energies by inward diffusion driven by Pc5 ULF waves with peak intensity in the 2--3.5 mHz frequency range.
\end{enumerate}

Our study demonstrates the critical importance of substorm injections to the outer radiation belt dynamics by showing that enhancements of ultra--relativistic electrons may occur even during weak magnetic storms, if chorus wave activity -- due to intense substorms -- is sufficiently intense. These chorus waves accelerated the seed population up to relativistic energies. In addition, the inward diffusion driven by the enhanced Pc5 ULF waves was able to further accelerate these electrons to ultra--relativistic energies up to $\approx10$ MeV creating an enhancement similar to the extreme storm of the St. Patrick's event of 2015 \citep{Li2016}, with $Sym-H_{min} \approx -235 nT$. 

Given the fact that the acceleration processes leading to the enhancement of ultra--relativistic electrons operate below geosynchronous orbit, the use of radiation monitors on--board MEO satellites would significantly advance our nowcasting and forecasting abilities with regard to highly relativistic electron enhancements.

% ------------------------------------------------------------------------ %
%  ACKNOWLEDGMENTS
\acknowledgments
The authors acknowledge the RBSP/MagEIS and RBSP/REPT teams for the use of the corresponding datasets. MagEIS (release 4) and REPT (release 3) data are available in https://www.rbsp-ect.lanl.gov/science/DataDirectories.php. Wen Li would like to acknowledge the NASA grant NNX17AD15G, the AFOSR grant FA9550-15-1-0158, the NSF grant AGS-1723588, and the Alfred P. Sloan Research Fellowship FG-2018-10936. We acknowledge the World Data Center for Geomagnetism, Kyoto for providing Kp and Dst indices and NASA CDAWeb for providing SYM-H index and solar wind parameters' data (https://omniweb.gsfc.nasa.gov/), Solar Influences Data Analysis Center for the Weekly bulletin on Solar and Geomagnetic activity, Issued: May 15, 2017 at 1451 UTC (http://sidc.be/archive) and the POES/SEM team for the use of the corresponding dataset (https://satdat.ngdc.noaa.gov/sem/poes/data/).
% ------------------------------------------------------------------------ %

\end{document}